\relax
\documentclass[letterpaper]{article} 
\usepackage{aaai22}  
\usepackage{times}  
\usepackage{helvet}  
\usepackage{courier}  
\usepackage[hyphens]{url}  
\usepackage{graphicx} 
\usepackage{natbib}  
\usepackage{caption} 
\DeclareCaptionStyle{ruled}{labelfont=normalfont,labelsep=colon,strut=off} 
\usepackage{algorithm}
\usepackage{algorithmic}
\usepackage{color}
\usepackage{amsmath}
\usepackage{graphicx}
\usepackage{booktabs} 

\usepackage{hyperref}
\hypersetup{
    colorlinks=true,
    citecolor=black,
    urlcolor=magenta,
}

\usepackage{newfloat}
\usepackage{listings}
\floatstyle{ruled}
\newfloat{listing}{tb}{lst}{}
\floatname{listing}{Listing}
\title{Triangular Character Animation Sampling with Motion, Emotion, and Relation}
\author {
    Yizhou Zhao\textsuperscript{\rm 1},
    Liang Qiu\textsuperscript{\rm 1},
    Wensi Ai\textsuperscript{\rm 2},
    Pan Lu\textsuperscript{\rm 1},
    Song-Chun Zhu\textsuperscript{\rm 1}
}
\affiliations {
    \textsuperscript{\rm 1}UCLA Center for Vision, Cognition, Learning, and Autonomy \\
    \textsuperscript{\rm 2}Department of Computer Science, UCLA \\
    yizhouzhao@g.ucla.edu
}

\usepackage{bibentry}

\begin{document}

\maketitle

\begin{abstract}
Dramatic progress has been made in animating individual characters. However, we still lack automatic control over activities between characters, especially those involving interactions. In this paper, we present a novel energy-based framework to sample and synthesize animations by associating the characters' body motions, facial expressions, and social relations. We propose a Spatial-Temporal And-Or graph (ST-AOG), a stochastic grammar model, to encode the contextual relationship between motion, emotion, and relation, forming a triangle in a conditional random field. We train our model from a labeled dataset of two-character interactions. Experiments demonstrate that our method can recognize the social relation between two characters and sample new scenes of vivid motion and emotion using Markov Chain Monte Carlo~(MCMC) given the social relation. Thus, our method can provide animators with an automatic way to generate 3D character animations, help synthesize interactions between Non-Player Characters (NPCs), and enhance machine emotion intelligence (EQ) in virtual reality (VR). 
\end{abstract}

\section{Introduction}
Traditional 3D animation is time-consuming. From sculpting 3D meshes, building rigging and deformation, to assigning skin weights, those preparatory steps make a 3D character ready to move. Beyond those efforts, an animator designs detailed movement for each body joint by setting keyframes or obtaining live actions from a motion capture device. That complicated work requires precise adjustment and careful consideration of many details.

\begin{figure*}[t]
    \centering
    \includegraphics[width=0.95\textwidth]{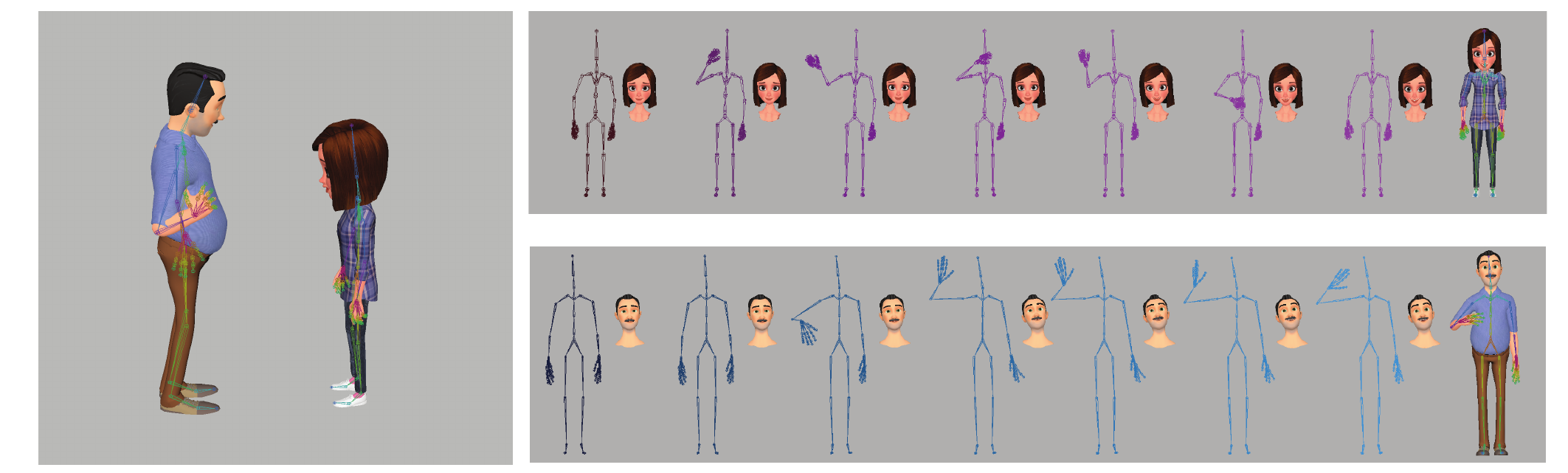}
    \caption{When animating a two-character social interaction scene, we need to consider the consistency of each character's body movements and facial expressions and the interplay between them. The figure on the left shows the initial frame of the animation; the figures on the right plot the sampling results of our method for the body movements and facial expressions of the two characters every $0.5$ second.}
    \label{figure:intro}
\end{figure*}

Recent years have witnessed the rapid evolution of machine learning methods that facilitate animation making~\cite{lee2018interactive, zhang2018data, taylor2017deep}. However, few works aim to make multi-character animation from a data-driven approach. In multi-character animations, movements of one character along with others, which bring out complicated combinations of body poses, hand gestures, and facial expressions. To synthesize meaningful animations, especially for multi-character social interactions, we argue there exist three main difficulties:
\begin{itemize}
\item Multi-character animation should interpret meaningful social interactions. (e.g., two characters say greetings with a high five.)
\item Body motions of one character must correspond with its facial emotions. For example, one character applauds with a happy face.
\item Multi-character animation brings additional constraints between characters to match their body movements and facial expressions temporally.
\end{itemize}
To address these challenges, first, we use the  Spatial-Temporal And-Or Graph (ST-AOG)~\cite{xiong2016robot} that samples a two-character key-frame animation from a stochastic grammar. The ST-AOG allows us to set up the contextual constraints for (body) motions, (facial) emotions, and the (social) relation between the two characters across time. Then, we propose Iterative Data Generating And Labeling (IDGAL) to alternatively sample scenes, label samples, and update model parameters, allowing us to efficiently train the ST-AOG from limited human-annotated data. Next, to make detailed animation for single-character motion and emotion, we apply the Eigenface method~\cite{turk1991eigenfaces} to encode facial joints for generating facial expressions, and a Variational Recurrent Neural Network (VRNN)~\cite{chung2015recurrent} to control body joints for generating body poses. Finally, using Markov Chain Monte Carlo (MCMC), the well-trained ST-AOG helps the Eigenface and VRNN to sample animations.

Our work makes two major contributions: (1) we are the pioneer to jointly and automatically sample the motion, emotion, and social relation for a multi-character configuration; (2) we present IDGAL to collect data while training a stochastic grammar. We plan to make our work (including our model and collected dataset) open-source to encourage researches on synthesizing multi-character animation.

The following sections first review some related works and then define the ST-AOG representing a two-character scene. Next, we formulate the stochastic grammar of the ST-AOG and propose the learning algorithm with IDGAL. Finally, after pre-training the Eigenface and VRNN, we can sample animations using MCMC~\footnote{The animation part in our work is done by Autodesk Maya~\cite{maya}}.

\section{Related Work}

\subsection*{Machine learning for animation}
Recent advances in machine learning have made it possible for animators to automatize some of the tough processes of generating human motion sequences. From early approaches such as hidden Markov models \cite{tanco2000realistic, ren2005data}, Gaussian processes \cite{wang2007gaussian, fan2011gaussian} and restricted Boltzmann machines \cite{taylor2009factored}, to more recent neural architectures such as Convolutional Neural Networks (CNN) \cite{holden2016deep} and Recurrent Neural Networks (RNN) \cite{fragkiadaki2015recurrent}, synthesizing human motions relies on framing poses or gestures~\cite{pavllo2019modeling}. More recent work focuses on improving animation quality by considering physical environment~\cite{holden2017phase}, character-scene interactions~\cite{starke2019neural} or social activities~\cite{shu2016learning}.

Emotion animation often comes together with speech animation \cite{taylor2017deep}. Producing high-quality speech animation relies on speech-driven audios~\cite{Oh_2019_CVPR} and performance-driven videos obtained by facial motion captures~\cite{maurer2001wavelet, cao2015real}. Then, animations for the designated actor can be synthesized by transferring facial features~\cite{taylor2017deep} or Generative Adversarial Networks (GAN)~\cite{vougioukas2019realistic}. 

\subsection*{Animation with social interaction}
Involving animations with social interactions has been studied for a long time \cite{takeuchi1995situated, waters1997diamond,arafa2002two, silvio2010animation}. Animations with social meanings are created by the interaction between the individuals in a group, which build better adoption for the characters in VR \cite{wei2019vr}, AR \cite{miller2019social}, and Human-robot Interaction (HRI) \cite{shu2016learning}.  Existing datasets for two-person interactions can be video-based \cite{yun2012two}, skeleton-based \cite{liu2019skeleton} or dialogue-based \cite{yu2020dialogue}. 


\subsection*{Stochastic grammar model}
Stochastic grammar models are useful for parsing the hierarchical structures for creating indoor scenes \cite{qi2018human}, predicting human activities \cite{taylor2017deep}, and controlling robots \cite{xiong2016robot}.
In this paper, we forward sampling from a grammar model to generate large variations of two-character animations. We also attempt to outline a general multi-character animation model to bring together motions, emotions, and social relations.

\begin{figure*}[t]
    \centering
    \includegraphics[width=0.75\textwidth]{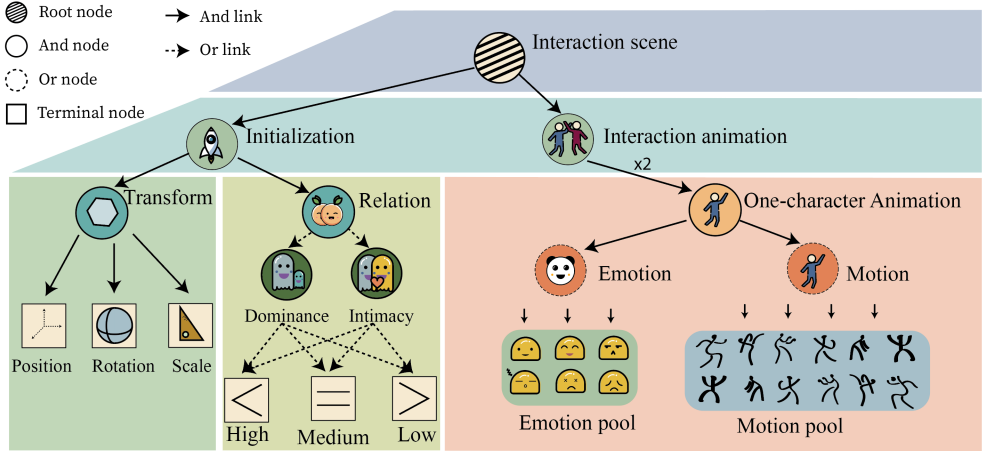}
    \caption{Scene grammar as an ST-AOG. We first initialize a social interaction scene between two characters by setting up the relative transform (position, rotation, and scale) and the type of the social relation between them. Then, for each character, we sample the animation that consists of one motion and one emotion.}
    \vspace{-3mm}
    \label{figure:aog}
\end{figure*}

\section{Representation and formulation}
In this section, we first deliver a brief mathematical definition of skeletal animation. Then, we introduce norms of \textbf{valence}, \textbf{arousal}, \textbf{dominance} and \textbf{intimacy}, which set constraints between motion, emotion, and social relation. Finally, we conduct the stochastic grammar for animations.

\subsection{Skeletal animation}
\textit{Skeletal animation} is a computer animation technique that enables a hierarchical set of body joints to control a character. Let $j$ denote one body joint that is characterized by its joint type, rotation, and position. A body pose $p$ is defined as a set of joints $\{j_u\}_{u=1,2,...,n}$ controlling the whole body, where $n$ is the total number of joints. Similarly, a facial expression $f$ is characterized by facial rigging. We can make an animation by designing the body pose $p$ and facial expression $f$ at $k$-th keyframe at time $t_k$. 

We define a \textbf{motion} $m$ as a sequence of body poses and an \textbf{emotion} $e$ as a sequence of facial expressions corresponding to the keyframes.
\subsection{Valance, arousal, dominance, and intimacy}
\begin{figure}[h]
    \centering
    \includegraphics[width=.45\textwidth]{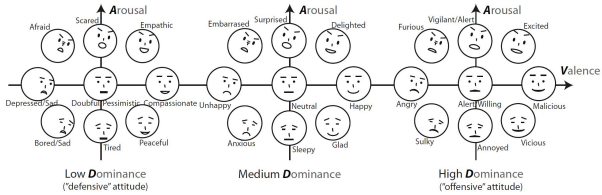}
    \caption{Norms of VAD and facial expressions~\cite{ahn2012nvc}}
    \vspace{-3mm}
    \label{figure:vad}
\end{figure}

Norms of valence, arousal, and dominance (VAD) are standardized to assess environmental perception, experience, and psychological responses~\cite{warriner2013norms}. \textbf{Valence} $v$ describes a stimulus's pleasantness, \textbf{arousal} $a$ quantifies the intensity provoked by a stimulus, and \textbf{dominance} $d$ evaluates the degree of control~\cite{bakker2014pleasure}. Norms of VAD can also describe a facial expression~\cite{ahn2012nvc}. Figure \ref{figure:vad} plots different facial expressions with varying degrees of valence, arousal, and dominance.

\begin{figure}[h]
    \centering
    \includegraphics[width=0.30\textwidth]{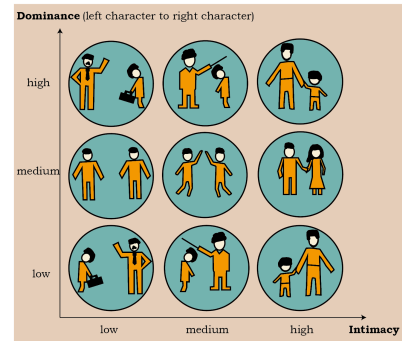}
    \caption{Examples of different social relations and their corresponding dominance-intimacy scores.}
    \vspace{-3mm}
    \label{figure:relation_types}
\end{figure}

Bringing the concept of \textit{intimacy}, we extend VAD norms as VADI norms. Specifically, \textbf{intimacy} $i$ describes the closeness of the relationship~\cite{karakurt2012relationship}. We describe the \textbf{social relation} $r$ between two characters by their relative dominance and intimacy $(d, i)$. Figure~\ref{figure:relation_types} plots different types social relations along with their dominance and intimacy.

Norms of valance, arousal, dominance, and intimacy form the space to set constraints between motion, emotion, and social relation. We will discuss it in detail in the next section.

\subsection{Stochastic grammar for two-character animations}
We define a Spatial-Temporal And-Or Graph (ST-AOG)
\begin{equation}
    \mathcal{G}=(R, V, C, P, S, T)
\end{equation}
to represent the social interaction scene of two characters, where $R$ is the root node for representing the scene, $V$ the node set, $C$ the set of production rules (see Figure \ref{figure:aog}), $P$ the probability model. The spatial relation set $S$ represents the contextual relations between terminal nodes and the temporal relation set $T$ represents the time dependencies.

\textbf{Node Set} $V$ can be decomposed into a finite set of nonterminal and terminal nodes: $V = V^{NT} \cup V^T$. The non-terminal nodes $V^{NT}$ consists of two subsets $V^{And}$ and $V^{Or}$. A set of \textbf{And-nodes} $V^{And}$ is a node set in which each node represents a decomposition of a larger entity (e.g., one character's animation) into smaller components (e.g., emotion and motion). A set of \textbf{Or-nodes}
$V^{Or}$ is a node set in which each node branches to alternative decompositions (e.g., the intimacy score between two characters can be low, medium, or high). The selection rule of Or-nodes follows the probability model $P$. In our work, we select the child nodes with equal. The \textbf{terminal nodes} $V^{T}$ represent entities with different meanings according to context: terminal nodes under the \textit{relation branch} identify the relationship between the two characters; the ones under \textit{motion branch} determine their motions, and the ones under \textit{emotion branch} depict emotions.

\textbf{Spatial Relations} $S$ among nodes are represented
by the horizontal links in ST-AOG forming Conditional Random Fields (CRFs) on the terminal nodes. We define different potential functions to encode pairwise constraints between motion $m$, emotion $e$, and social relation $r$:
\begin{equation}
S = S_{me}\cup S_{re} \cup S_{rm} .
\end{equation}
$S_{me}$ sets constraints on the motion and emotion to ensure that the body movement supports one emotion according to the social affordance. For example, the crying action (rubbing eyes) can hardly be compatible with a big smile. $S_{re}$ regulates the emotion when social relation is considered. For instance, an employee has few chances to laugh presumptuously in front of the boss. Similarly, $S_{rm}$ manages to select the suitable body motion under social relation. For example, kissing is allowed for couples. 


\textbf{Temporal Relations} $T$ among nodes are also represented by the links in ST-AOG to address time dependencies. Temporal relations are divided into two subsets:
\begin{equation}
T = T_{me} \cup T_{r} .
\end{equation}
$T_{me}$ encodes the temporal associations between motion and emotion. $T_{r}$ describes the extent to which the two characters' animations match temporally.

A hierarchical parse tree $pt$ is an instantiation of the ST-AOG by selecting a child node for the Or-nodes and determining the terminal nodes. A parse graph $pg$ consists of a parse tree $pt$, a number of spatial relations $S$, and several temporal relations $T$ on the parse tree:
\begin{equation}
    pg = (pt, S_{pt}, T_{pt}) .
\end{equation}

\subsection{Probabilistic model of ST-AOG}
A scene configuration is represented by a parse graph $pg$, including animations and social relations of the two characters. The probability of $pg$ generated by an ST-AOG parameterized by $\theta$ is formulated as a Gibbs distribution:
\begin{small}
\begin{align} \label{eq:energy}
p(pg \mid \Theta) &=\frac{1}{Z} \exp \{-\mathcal{E}(p g \mid \Theta)\} \nonumber\\
&=\frac{1}{Z} \exp \left\{-\mathcal{E}(p t \mid \Theta)-\mathcal{E}\left(S_{p t} \mid \Theta\right) -\mathcal{E}\left(T_{p t} \mid \Theta\right) \right\} ,
\end{align}
\end{small}
\noindent where $\mathcal{E}(p g \mid \Theta)$ is the energy function of a parse graph, and $\mathcal{E}(p t \mid \Theta)$ of a parse tree. $\mathcal{E}(S_{p t} \mid \Theta)$ and $\mathcal{E}(T_{p t} \mid \Theta)$ are the energy terms of spatial and temporal relations.
$\mathcal{E}(pt \mid \Theta)$ can be further decomposed into the energy functions of different types nodes:
\begin{equation}
\mathcal{E}(p t \mid \Theta)=\underbrace{\sum_{v \in V} \mathcal{E}_{\Theta}^{O r}(v)}_{\text {non-terminal nodes }}+\underbrace{\sum_{v \in V_{T}^{r}} \mathcal{E}_{\Theta}^{T}(v)}_{\text {terminal nodes }} .
\end{equation}

\textbf{Spatial potential} $\mathcal{E}\left(S_{p t} \mid \Theta\right)$ combines the potentials of three types of cliques $C_{me},C_{re}, C_{rm}$ in the terminal layer, integrating semantic contexts mentioned previously for motion, emotion and relation:
\begin{align}
p\left(S_{p t} \mid \Theta\right) &=\frac{1}{Z} \exp \left\{-\mathcal{E}\left(S_{p t} \mid \Theta\right)\right\} \nonumber\\
&=\prod_{c \in C_{me}} \phi_{me}(c) \prod_{c \in C_{re}} \phi_{re}(c) \prod_{c \in C_{rm}} \phi_{rm}(c) .
\end{align}
We apply the norms of valence, arousal, dominance and intimacy (VADI) to quantify the triangular constraints between social relation, emotion and motion:  
\begin{itemize}
    \item Social relation $r$ is characterized by its dominance and intimacy $(d_r, i_r)$.
    \item For emotion $e$, which is a sequence facial expression, we consider its valance, arousal, and dominance scores as the different between the beginning facial expression $f_0$ and ending facial expression $f_1$:
    \begin{equation}
        (v_e, a_e, d_e) = (v_{f_1}, a_{f_1}, d_{f_1}) - (v_{f_0}, a_{f_0}, d_{f_0}) .
    \end{equation}
    \item To get the scores of a motion $m$, we first label the name $N_m$ of the motion, such as \textit{talk}, \textit{jump} and \textit{cry}. Then we can obtain the valance, arousal, and dominance scores from NRC-VAD Lexicon \cite{mohammad2018obtaining}, which includes a list of more than 20,000 English words and their valence, arousal, and dominance scores:
    \begin{equation*}
        m \to N_m \to (v_m, a_m, d_m) .
    \end{equation*}
\end{itemize}

Therefore, the relation $S_{me}$ and its potential $\phi_{me}$ on the clique $C_{me} = \{(m, e)\}$ contain all the motion-emotion pairs in the animation, and we define
\begin{align}\label{eq:S_me}
     \phi_{me}(c) = \frac{1}{Z_{me}^s}\exp\{\lambda_{me}^s\cdot (v_m, a_m, d_m) \cdot (v_e, a_e, d_e)^\top\}.
\end{align}

Calculating potentials $\phi_{rm}$ on clique $C_{rm} = \{(m, r)\}$ and $\phi_{re}$ on $C_{re} = \{(e, r)\}$ needs another variable $i_{me}$ suggesting the intimacy score. $i_{me}$ is defined as the distance $dist$ between the two characters compared with a standard social distance $dist_0$ :
\begin{equation}
    i_{me} = (dist_0 - dist) / dist_0 .
\end{equation}
Then we can define
\begin{align}
    \phi_{re}(c) &= \frac{1}{Z_{re}^s}\exp\{\lambda_{re}^s\cdot (d_r, i_{r}) \cdot (d_e, i_{me})^\top\} , \label{eq:S_re} \\
    \phi_{rm}(c) &= \frac{1}{Z_{rm}^s}\exp\{\lambda_{rm}^s\cdot (d_r, i_{r}) \cdot (d_m, i_{me})^\top\} . \label{eq:S_rm}
\end{align}

\textbf{Temporal potential} $\mathcal{E}\left(S_{p t} \mid \Theta\right)$ combines two potentials for time control, and we have 
\begin{align}
p\left(T_{p t} \mid \Theta\right) &=\frac{1}{Z} \exp \left\{-\mathcal{E}\left(T_{p t} \mid \Theta\right)\right\} \nonumber \\
&=\prod_{c \in C^T_{me}} \psi_{me}(c) \prod_{c \in C^T_{r}} \psi_{r}(c) .
\end{align}
Potential $\psi_{me}$ is defined on clique $C^T_{me} = \{(t_{m}, t_{e})\}$ representing the time to start a motion and an emotion. We assume the time discrepancy between them follows a Gaussian distribution, then we can get 
\begin{align}\label{eq: T_me}
    \psi_{me}(c) =  \frac{1}{Z^t_{me}}\exp \left(\lambda_{re}^t \cdot (t_{m} - t_{e})^2\right) .
\end{align}
Notice that so far the training parameters $\lambda_{me}^s$, $\lambda_{rm}^s$, $\lambda_{rm}^s$, $\lambda_{me}^t$ and partition functions $Z_{me}^s$, $Z_{re}^s$, $Z_{rm}^s$, $Z_{me}^t$ should be doubled since we have two characters in the scene.

At last, to match the animation for both characters, we assume that the time differences between ending time of their motions $t_{1,m}, t_{2,m}$ and emotions $t_{1,e}, t_{2, e}$ follow the Gaussian distribution: 
\begin{align}\label{eq: T_r}
    \psi_{r}(c) &= \frac{1}{Z^t_{m}}\exp \left(\lambda_{m}^t \cdot (t_{1,m} - t_{2,m})^2\right) \nonumber\\ 
    & + \frac{1}{Z^t_{e}}\exp \left(\lambda_{e}^t \cdot (t_{1,e} - t_{2,e})^2\right) .
\end{align}
Here we have two additional parameters $\lambda_{m}^t, \lambda_{e}^t$, and two more partition functions $Z_{m}^t, Z_{e}^t$.

\section{Data collection}
Traditional methods \cite{creswell2018generative} for training a generative model prefer using a well-labeled dataset. Therefore, a model hardly modifies the original dataset after preparation steps such as data preprocessing and augmentation. However, there is not a suitable dataset to train our model. Fortunately, our ST-AOG is a generative model that can sample animations.

We start from low-quality animations sampled by our ST-AOG, since our ST-AOG is initiated from a random state and probably sample distorted body poses, weird facial expressions, and meaningless interpretations for the two-character social interactions. Then, we introduce \textbf{iterative data generating and labeling} (IDGAL) to render two-character interaction scenes, making more samples while improving the data sampling quality by training model. Each round of IDGAL contains three steps: sampling scenes from our ST-AOG, labeling scenes, and updating ST-AOG using the labeled data.

\begin{figure}[h]
    \centering
    \includegraphics[width=0.35\textwidth]{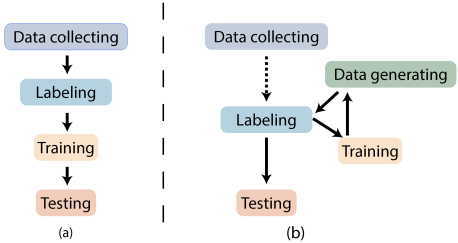}
    \vspace{-2mm}
    \caption{Comparison between  (a) regular machine learning pipeline and (b) iterative data generating and labeling.}
    \vspace{-2mm}
    \label{fig:my_label}
\end{figure}

\subsection{Sampling scenes}
Sampling $pg$ from our ST-AOG requires selecting Or-nodes and terminal nodes from the \textit{transform branch}, \textit{relation branch}, \textit{emotion branch}, and \textit{motion branch}. We assume that each character in a scene has one motion, one starting facial expression and one ending facial expression.

The \textit{transform branch} samples the initial relative distance between the two characters and we assume that they stand face to face. The \textit{relation branch} sets up their social relation (see Figure \ref{figure:relation_types}). The ST-AOG samples emotions from a pool of $21$ basic facial expressions (e.g. smile, happy face, and sad face) by facial rigging from Facial Action Coding System \cite{cohn2007observer} and annotate their VAD scores by the NRC-VAD Lexicon~\cite{mohammad2018obtaining}. The motion pool we gathered is from Adobe Mixamo \cite{blackman2014rigging}. We select a total number of $65$ single-person animations and label their names (such as \textit{kissing}, \textit{bowing} and \textit{yelling}) in order to get their VAD scores by the NRC-VAD Lexicon~\cite{mohammad2018obtaining}. 

The above process specifies the configuration of $pg$ and VADI scores that determine the spatial potential $\mathcal{E}\left(S_{p t} \mid \Theta\right)$. To obtain temporal potential $\mathcal{E}\left(T_{p t} \mid \Theta\right)$, We further assume the time difference between motion start and emotion start, and the time misalignments between their motions and emotions $(t_{1,m} - t_{2,m})$ and $(t_{1,e} - t_{2,e})$ follow the standard normal distribution. 


\subsection{Labeling scenes}
Training our ST-AOG requires labeling expert data (see Equation \ref{eq:train} and \ref{eq:gradient}): the model learns to sample those expert data with higher probabilities. Therefore, we label the scene's quality by asking experimenters' subjective judgments on whether the scene is reasonable and meaningful. 

\subsection{Updating ST-AOG}
We rewrite the probability
distribution of cliques formed on terminal nodes as 
\begin{align} \label{eq:train}
p\left(S_{p t}, T_{p t}, \mid \Theta\right) &=\frac{1}{Z} \exp \left\{-\mathcal{E}\left(S_{p t} \mid \Theta\right) -\mathcal{E}\left(T_{p t} \mid \Theta\right)  \right\} \nonumber \\ &=\frac{1}{Z} \exp \left\{-\left\langle\lambda, l\left(ST_{p t}\right)\right\rangle\right\} .
\end{align}
where $\lambda$ is the weight vector and $l\left(ST_{p t}\right)$ is the loss vector given by Equations \ref{eq:S_me}, \ref{eq:S_re}, \ref{eq:S_rm}, \ref{eq: T_me} and \ref{eq: T_r}.
To learn the weight vector, the standard maximum likelihood estimation (MLE) maximizes
the log-likelihood:
\begin{equation}
\mathcal{L}\left(S_{p t}, T_{p t}, \mid \Theta\right)=-\frac{1}{N} \sum_{n=1}^{N}\left\langle\lambda, l\left(ST_{pt} \right)\right\rangle-\log Z .
\end{equation}
It is usually maximized by gradient descent:
\begin{align}\label{eq:gradient}
&\frac{\partial \mathcal{L}\left(ST_{p t} \mid \Theta\right)}{\partial \lambda}=-\frac{1}{N} \sum_{n=1}^{N} l\left(ST_{p t_{n}}\right)-\frac{\partial \log Z}{\partial \lambda} \nonumber \\
&=-\frac{1}{N} \sum_{n=1}^{N} l\left(ST_{p t_{n}}\right)+\frac{1}{\tilde{N}} \sum_{\tilde{n}=1}^{\tilde{N}} l\left(ST_{p t_{\tilde{n}}}\right) ,
\end{align}
where $\{l(ST_{p t_{\tilde{n}}})\}_{\widetilde{n}=1, \cdots, \widetilde{N}}$ is calculated by synthesized examples from the current model, and $\{l(ST_{p t_{n}})\}_{n=1, \cdots, N}$ by expert samples (gold labels) from the labeled dataset. 


\subsection{A dataset of two-character animations}
We apply Equation \ref{eq:gradient} to train our model from the labeled data.  We truncate the low-likelihood samples, and in each updating ST-AOG round, we take $100$ epochs of the gradient descent (equation \ref{eq:gradient}) with the learning rate as $1e-3$. After getting the sampled scenes in each round, experimenters were asked to examine scenes and label them with good/medium/bad. In total, three rounds of IDGAL give us $1,240$ well-labeled animations with a total length of $9,285$ seconds. Table \ref{table:iterative} shows the improvement along with three rounds of IDGAL: the percent of good samples rises whereas the percent of bad samples decreases. 

\begin{table}[h]
\begin{tabular}{c||ccc}
\hline
          & \textbf{Round one} & \textbf{Round two} & \textbf{Round three} \\ \hline\hline
\textbf{Count}     & 440     & 400     & 400     \\
\textbf{Good rate} & 36.8\%  & 39.5\%  & 40.0\%  \\
\textbf{Bad rate}  & 36. 0\% & 26.5\%  & 24.0\%  \\ \hline
\end{tabular}
\caption{Quality labeling results of different rounds.}
\label{table:iterative}
\end{table}

\section{Synthesizing scenes}
Training our ST-AOG requires sampling motions and emotions from existing motion and emotion pools, limiting it to a small range of applications with little flexibility. To overcome this problem, we present the idea to make the learned stochastic grammar suitable for all kinds of generative models for facial animation and body movement so that our ST-AOG can be useful for a broader range of applications. 

\subsection{Markov chain dynamics}

First, we design three types of Markov chain dynamics: (1)~$q_{r}$ to make proposal sample social relations;
(2) $q_e$ to sample emotions; (3) $q_m$ to sample motions.

\textbf{Relation dynamic} $q_r$ makes transition of social relations directly from ST-AOG's Or-nodes on \textit{relation branch}: 
\begin{align}
\text{Node}_{r_1}\to\text{Node}_{r_2} .
\end{align}

\textbf{Emotion dynamic} $q_e$ changes the VAD scores of one of the facial expression at keyframe $t_k$:
\begin{align}
(v,a,d) \to (v^\prime,a^\prime,d^\prime) .
\end{align}
Besides, to synthesize emotions, we train a linear model based on our $21$ basic faces with manually labeled VAD scores. Specifically, we first get the eigenfaces through principal component analysis (PCA) based on the positions of \textit{eyebrows, eyes, mouth} and \textit{cheeks}. We show some constructed examples in Figure \ref{figure:reconstructed_emotion} and leave the detailed analysis in the appendix.

\begin{figure}[h]
    \centering
    \includegraphics[width=0.85\linewidth]{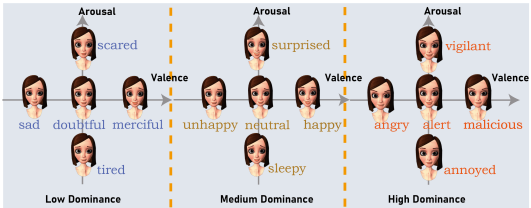}
    \caption{Reconstructed facial expressions and VAD scores.}
    \label{figure:reconstructed_emotion}
\end{figure}

\textbf{Motion dynamic $q_m$} regards motion as a sequence of body poses $\{p_i\}_{i=1,2,...}$.  Therefore, we train a generative model for motions, where the model takes the inputs of poses $\{p_i\}_{i=1,2,...,k}$ for every $\delta t$ time interval and predicts the next body pose $p_{k+1}$ according to the maximum likelihood:
\begin{align}
    \arg\max_{p_{k+1}}\text{P}(p_{k+1}\mid p_{k}, p_{k-1}, ..., p_{1}; \delta t) .
\end{align}

We train a model to generate animations for a single character. The motion samples collected from Adobe Mixamo~\cite{blackman2014rigging} are manually filtered. The filtered database contains $149$ animations such as \textit{yelling, waving a hand, shrugging}, and \textit{talking}. The complete list of animations is attached to the Appendix. 

The data augmentation process mirrors each animation from left to right, resulting in a total number of $36,622$ keyframes of animations with $24$ frames in a second. We set $0.5$s as the time interval between poses, and each pose is represented by the rotations of $65$ joints and the position of the root joint (see Appendix for the rigging). Therefore, each $p_t$ is a $198$-dimensional ($65\times 3 + 3$) vector. 
 
\begin{figure}[h]
    \centering
    \includegraphics[width=0.65\linewidth]{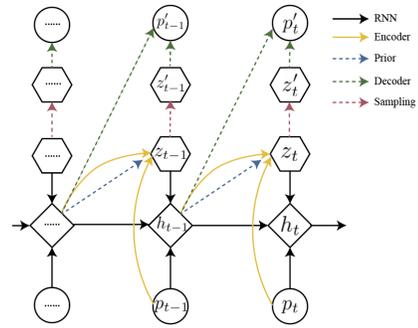}
    \caption{The variational recurrent neural network (VRNN).}
    \label{figure:vrnn}
\end{figure}

 We apply the variational recurrent neural network (VRNN) \cite{chung2015recurrent} as our motion generative model and we set $\delta t = 0.5\text{s}$. As Figure \ref{figure:vrnn} shows, the model first encodes the historical poses $\{p_{u}\}_{u=1,2,...,{k-1}}$ into a state variable $h_{k-1}$, whereby the prior on the latent random variable $z_{k}$ is calculated. $z_{k}$ usually follows a Gaussian distribution. Finally, the generating process takes the inputs $h_{k-1}$ and $z_{k}$ to generate pose $p_k$. 
 
The training procedure follows that of a standard VAE \cite{doersch2016tutorial}. The objective is to minimize the reconstruction loss (mean square error between original and reconstructed data) and KL-divergence loss. The results show that VRNN performs well: $56.9\%$ of the reconstructed errors for joint rotations are less than $1.0$ degree and $87.3\%$ less than $5$ degrees. See to the Aappendix for detailed analysis.

Then, dynamic $q_m$ generates a body pose $p_k$ by sampling the latent random variable $z_k$:
\begin{align}
    &p_{k} \to p_{k}^\prime , \\ 
    \text{ by }  &z_{k} \to z_{k}^\prime .
\end{align}
To map the pose to the VADI norms, we train a linear model:
\begin{align}
    p_k \xrightarrow{\text{regression}} (v,a,d,i) .
\end{align}

Finally, according to Equations \eqref{eq:S_me}, \eqref{eq:S_re}, and \eqref{eq:S_rm}, we can calculate the probability of sampling $pg^\prime$.

\begin{table*}[h]
\resizebox{0.98\textwidth}{!}{
\begin{tabular}{cccc}
\hline
\textbf{Scenario} & \textbf{Social relation} & \textbf{Character one emotion} & \textbf{Character two emotion}\\ \hline
wave hands                 & friends (medium, medium)   & neutral $\to$ happy            & \textcolor{blue}{$(0.5, 0.5, 0.5)$[neutral]} $\to$\textcolor{red}{$(0.9, 0.6, 0.6)[delight]$}                             \\
high-five                 & brothers (medium, high)   & happy $\to$ excited            & \textcolor{blue}{$(0.5, 0.5, 0.5)$[neutral]} $\to$\textcolor{red}{$(0.9, 0.7, 0.7)[glad]$}     \\
shake hands                 & strangers (medium, low)   & neutral $\to$ excited            & \textcolor{blue}{$(0.8, 0.7, 0.6)$[joyful]} $\to$\textcolor{red}{$(0.9, 0.4, 0.8)[respectful]$}    \\
apologize                & employee to employer (low, medium)   & neutral $\to$ sad            & \textcolor{blue}{$(0.5, 0.5, 0.5)$[neutral]} $\to$\textcolor{red}{$(0.3, 0.6, 0.4)[concerned]$}                             \\
criticize                 & teacher to student (high, close)   & neutral $\to$ angry            & \textcolor{blue}{$(1.0, 0.7, 0.8)$[happy]} $\to$\textcolor{red}{$(0.3, 0.8, 0.3)[scared]$}     \\
quarrel        & colleagues (medium, medium)   & neutral $\to$ dissatisfied            & \textcolor{blue}{$(0.5, 0.5, 0.5)$[neutral]} $\to$\textcolor{red}{$(0.1, 0.8, 0.3)[annoyed]$}    \\
\hline
\\
\end{tabular}
}
\caption{Emotion sampling results. The \textit{social relation} is labeled with its dominance and intimacy scores. MCMC generates the emotion to sample from the starting facial expression (the blue text describes its valence, arousal, and dominance scores) to the ending facial expression (VAD norms in red). We also pick a word to describe a facial expression according to the NRC-VAD Lexicon~\cite{warriner2013norms}.}
\label{Table:emotion}
\end{table*}

\subsection{MCMC}
Adopting the Metropolis-Hastings algorithm \cite{chib1995understanding}, the proposed new parse graph $pg\prime$ is sampled according to the following acceptance probability:
\begin{align}
\alpha\left(p g^{\prime} \mid p g, \Theta\right) &=\min \left(1, \frac{p\left(p g^{\prime} \mid \Theta\right) p\left(p g \mid p g^{\prime}\right)}{p(p g \mid \Theta) p\left(p g^{\prime} \mid p g\right)}\right), 
\end{align}
where $p\left(p g\mid \Theta\right)$ and $p\left(p g^{\prime}  \mid \Theta\right)$ are calculated by energy terms $\exp (\mathcal{E}(p g \mid \Theta))$ and $\exp (\mathcal{E}(p g^{\prime} \mid \Theta))$ from Equation \eqref{eq:energy}. $p(p g \mid p g^\prime)$ and $p(p g^\prime \mid p g)$ are the proposal probabilities define by dynamics $q_r, q_e$, and $q_m$.

\section{Application}
We present two applications of our model:
\begin{itemize}
    \item \textbf{Emotion sampling}: given two characters' social relations and motions, our model can complement animations of facials expressions.
    \item \textbf{Motion completion}: given two characters' social relation, emotions, and part of the movement information, our model can complement the subsequent body movements of characters.
\end{itemize}



\subsection{Emotion sampling}
Facial expressions are among universal forms of body language and nonverbal communication. Currently, there is a lack of study on how a virtual character could interplay his/her facial expression to another character, resulting in poor playability in a game or a virtual reality (VR) platform. For example, a user is shaking hands with a VR character, who offers his/her hand but does not show any facial expressions. His/Her expressionless face may make the user confused or even scared.

Our work proposes a method to automatically generate facial expressions for virtual characters. Given two characters' motions, their social relation, and the emotion of one of the characters, our task is to sample the ending facial expression of the other character from his/her starting facial expression. We consider six scenarios: \textit{wave hands} (between friends), \textit{high-five} (between brothers), \textit{shake hands} (between strangers), \textit{apologize} (employee to employer), \textit{criticize} (teacher to student), and \textit{quarrel} (between colleagues). Figure \ref{figure:ms_scene} depicts the snapshots of motions of two characters.

We apply the emotion dynamic $q_e$ for a total number of $20$ times for each scenario, and change the value $v$, $a$, or $d$ by $\pm 0.1$ for each proposed $pg^\prime$. Table \ref{Table:emotion} shows the sampling results. With the help of VAD scores, facial expression can be synthesized accurately.



\subsection{Motion completion}
Common methods to control the motions of a character including finite state machines and behavior trees highly rely on manually created animations. 
Our model also provides an innovative way to help generate body poses for characters. Given the two characters' emotions and their social relation, our task is to sample the other character's next body pose based on his/her previous poses. We consider the same six scenarios as those in emotion sampling. However, in this case, we only keep the motions of the left character and set up the starting pose of the right one as \textit{standing} in those scenarios. The social relation and emotions follow those in Table \ref{Table:emotion}. 

\begin{figure}[h]
    \centering
    \includegraphics[width=1.0\linewidth]{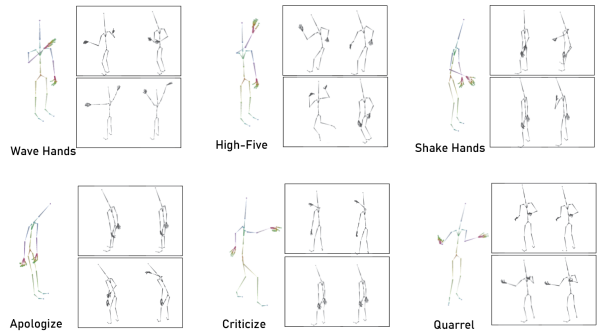}
    \caption{Motion sampling results. Our model samples the animations (shown in the rectangle) that represent the response of a character in different scenarios. For each scenario, we sample two sets of body poses.}
    \label{figure:ms_scene}
\end{figure}

Figure \ref{figure:ms_scene} plots the skeletons of sampled poses. For each scenario, we sample two pose sequences by VRNN that represent the character's body movement after $0.5$ second and $1.0$ second. We can see that in most cases, our sampled motions can interpret some social interaction meanings. For example, to respond to the left character's high-five proposal, the right character may raise his hand or jump happily.

\section{Conclusion}

We propose a method to generate animations of two-character social interaction scenes automatically. Our approach can be useful for the tasks including but not limited to: i) assisting animators to make keyframe animations including character poses and facial expressions. ii) helping game developers generate vivid NPC interaction events; iii) offering better emotional intelligence for VR agents.

\newpage

\bibliography{sample-bibliography}

\newpage

\end{document}